\documentclass[aps,prb,amsmath,amssymb,reprint,floatfix]{revtex4-1}

\usepackage{hyperref}
\usepackage{epsfig}
\usepackage{graphicx}
\usepackage{subfigure}
\usepackage{latexsym}
\usepackage{color}
\usepackage{fullpage}
\usepackage{dcolumn}
\usepackage{bm}
\usepackage{ulem}
\usepackage{units}

\begin{document}

\title{Proposal to recover an extensive ground state degeneracy in a two-dimensional square array of nanomagnets}

\author{Yann Perrin, Benjamin Canals and Nicolas Rougemaille}

\affiliation{Univ. Grenoble Alpes, CNRS, Grenoble INP, Institut NEEL, 38000 Grenoble, France}

\date{\today}

\begin{abstract}

We investigate numerically the micromagnetic properties and the low-energy physics of an artificial square spin system in which the nanomagnets are physically connected at the lattice vertices. 
Micromagnetic simulations reveal that the energy stored at the vertex sites strongly depends on the type of magnetic domain wall formed by the four connected nanomagnets. 
As a consequence, the energy gap between the vertex types can be partially modified by varying the geometrical parameters of the nanomagnets, such as their width and thickness. 
Based on the energy levels given by the micromagnetic simulations, we compute the thermodynamic properties of the corresponding spin models using Monte Carlo simulations. 
We found two regimes, both being characterized by an extensive ground state manifold, in sharp contrast with similar lattices with disconnected nanomagnets. 
For narrow and thin nanomagnets, low-energy spin configurations consist of independent ferromagnetic straight lines crossing the whole lattice. 
The ground state manifold is thus highly degenerate, although this degeneracy is subdominant. 
In the limit of thick and wide nanomagnets, our findings suggest that the celebrated square ice model may be fabricated experimentally from a simple square lattice of connected elements. 
These results show that the micromagnetic nature of artificial spin systems involves another degree of freedom that can be finely tuned to explore strongly correlated disordered magnetic states of matter.

\end{abstract}


\maketitle

\section{Introduction}

Artificial arrays of interacting nanomagnets were introduced \cite{Tanaka2005, Tanaka2006, Wang2006} as a mean to fabricate experimentally various types of spin and vertex models. 
In particular, the idea of using lithographically-patterned architectures to explore the physics of highly frustrated magnets \cite{HFM2011} triggered a wealth of studies at the frontier between nanomagnetism and condensed matter magnetism.
Classical spin liquids \cite{Qi2008, Sendetskyi2016, Rougemaille2011}, emerging magnetic properties \cite{Rougemaille2011, Moller2009, Zhang2013, Chioar2014-1, Montaigne2014, Brooks2014, Canals2016}, Coulomb phases \cite{Perrin2016, Ostman2018} and complex magnetic ordering \cite{Chioar2014-2, Chioar2016, Hamp2018} are examples of the low-energy physics that can be now probed experimentally, in an almost routine fashion.
Because almost any type of two-dimensional (2d) geometry can be designed, whether or not this geometry exist in nature \cite{Morrison2013, Gilbert2014, Gilbert2015, Stopfel2018, Shi2018}, artificial spin systems offer a powerful lab-on-chip approach to directly visualize exotic magnetic phenomena in real space \cite{Nisoli2013, Rougemaille2019}.

\begin{figure}
\center
\includegraphics[width=7.5cm]{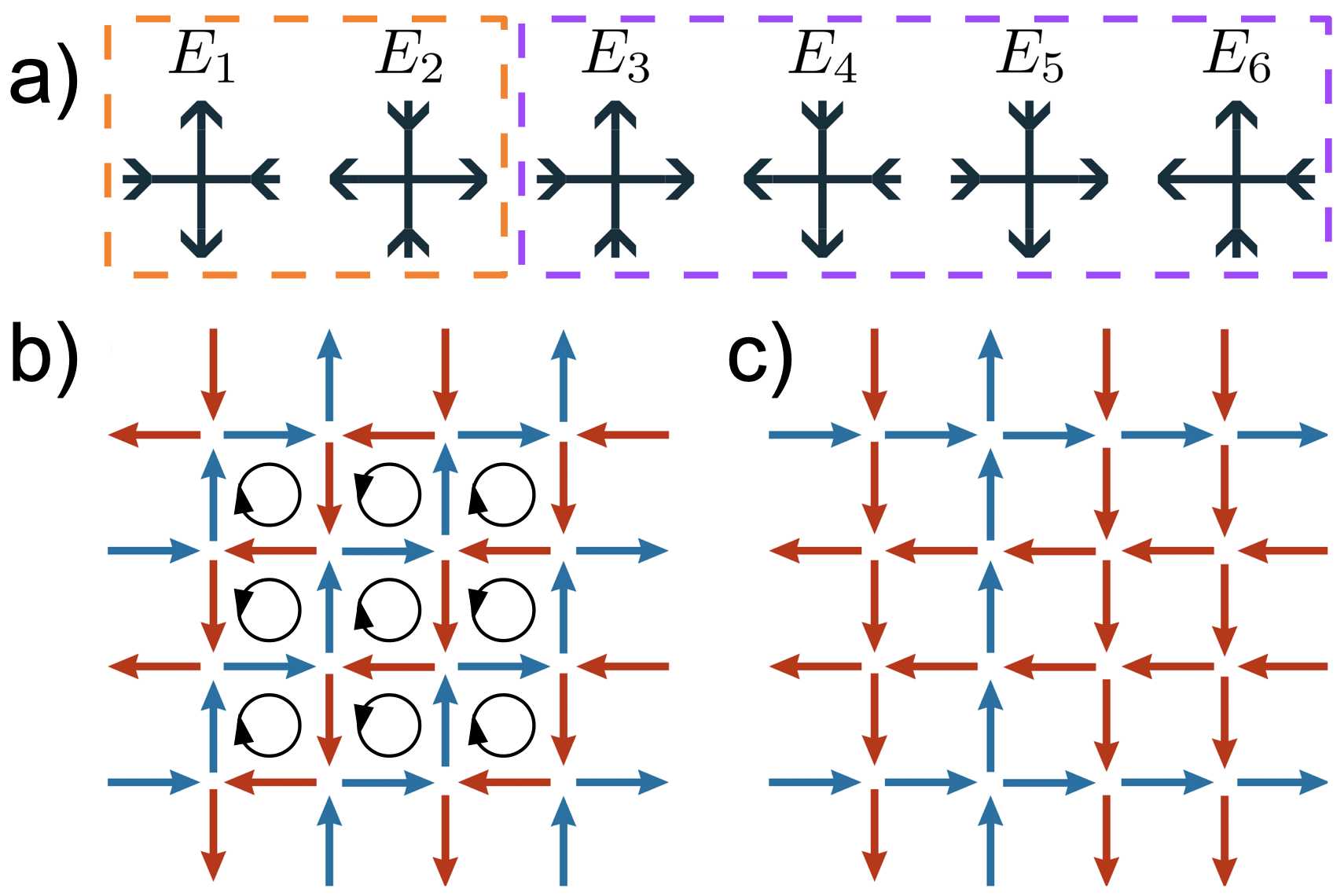}
\caption{\label{vertex} (a) Schematics of the vertices involved in the six vertex model with their associated energy $E$. Because of symmetry, these vertices can be sorted in two types only (see dashed rectangles), with $E_1=E_2=E_I$ and $E_3=E_4=E_5=E_6=E_{II}$ (b). Ground state configuration when $E_1<E_{II}$. (c) One given configuration belonging to the degenerate ground state manifold of the KDP model obtained when $E_1>E_{II}$.}
\end{figure}

Among the works done so far, the square geometry has been extensively studied \cite{Wang2006, Moller2006, Nisoli2010, Morgan2011, Phatak2011, Budrikis2011, Budrikis2012, Farhan2013, Porro2013, Kapaklis2014, Thonig2014}. 
One reason is that the square geometry potentially allows investigations of ice-type models, a family of vertex models introduced in the thirties by Linus Pauling to describe the residual entropy of water ice at low temperature \cite{Pauling1935}. 
However, due to the nonequivalent strength of the magnetostatic interaction between orthogonal and collinear elements in artificial 2d square arrays of disconnected nanomagnets, these systems were not able to reach the square ice physics.
Instead, these arrays order in a N\'{e}el-like fashion at low effective temperature, and the macroscopic degeneracy of the ground state manifold initially sought is lost, unless the 2d geometry is modified \cite{Perrin2016, Ostman2018}.

In this study, we show numerically that the extensive ground state manifold of the square-ice model can be observed in simple 2d magnetic structures.
Do to so, we consider square arrays of nanomagnets that extend up to the vertex, meaning that the nanomagnets are physically connected at the nodes of the lattice.
Connected square spin systems are reminiscent of several other works done in the past on magnetic antidots arrays \cite{Yu2000, Vavassori2002, Wang2003, Guedes2003, Heyderman2003}. 
Besides the magnetostatic interaction, micromagnetic exchange now plays a key role. 
Competition between these two interactions generates different micromagnetic textures \cite{Rougemaille2013, Burn2014, Burn2015, Walton2015, Gliga2015} depending on the magnetic orientations of the nanomagnets defining the vertex type. 
The energy associated with the magnetization distribution at a given vertex is estimated using micromagnetic simulations. 
Results show that the energy gap between the vertex types can be partly modified when varying the geometrical parameters of the nanomagnets, thus allowing to explore two variants of the six vertex models: the square ice and the KDP model.
Based on the energy levels given by micromagnetic simulations, we examine the thermodynamic properties of these two variants by Monte Carlo simulations.

\section{Artificial realization of the six vertex model}

In the six vertex model, each configuration is defined by the state of four arrows (Ising spins) located on the bonds of a square lattice.
Among the $2^4=16$ possibilities to define a vertex, only the six states made of two spins pointing inwards and two spins pointing outwards the vertex are considered in this model. 
In other words, only vertices having a divergence-free state are taken into account, while states being a source or sink of effective magnetic charge are not considered.
These six states are represented in Fig. \ref{vertex}a and can have six different energies.
Because of symmetry, in artificial arrays of nanomagnets, the conditions $E_1=E_2=E_I$ and $E_3=E_4=E_5=E_6=E_{II}$ are verified, such that the six vertices can be sorted in two groups [see Fig. \ref{vertex}a]: type-I vertices, with no net magnetic moment, and type-II vertices with a nonzero magnetic moment. 

As mentioned above, in artificial 2d arrays of disconnected nanomagnets, magnetostatics leads to the condition $E_I < E_{II}$. 
This property has an immediate consequence: the low-energy physics of these arrays is described by the F-model \cite{Rys1963, Lieb1967a}: the magnetic ground state is ordered and consists of flux-closure loops with alternating chirality [see Fig. \ref{vertex}b].
Artificial spin systems can be also designed in such a way that the condition $E_I = E_{II}$ is fulfilled. 
In that case, the low-energy physics of the array is described by the square ice model \cite{Lieb1967b} and the ground state manifold is macroscopically degenerate.
The magnetic system is an algebraic spin liquid, i.e., a spin liquid characterized by spin-spin correlations decaying with a power law with the separating distance \cite{Perrin2016}.
Local configurations breaking the divergence-free constraint associated with the six vertex model behave, in that particular case, and in that particular case only, as deconfined particule-like excitations interacting via a Coulomb potential \cite{Castelnovo2004, Henley2010}.

In the following, we consider the third possible situation of the six vertex model in which $E_I > E_{II}$.
The low-energy physics of the system is then described by the KDP model, \cite{Slater1941, Lieb1967c} initially introduced to describe ferroelectrics.
The ground state configuration is then made of decoupled ferromagnetic straight lines crossing the entire lattice [see Fig. \ref{vertex}c].

\begin{figure}
\center
\includegraphics[width=7.5cm]{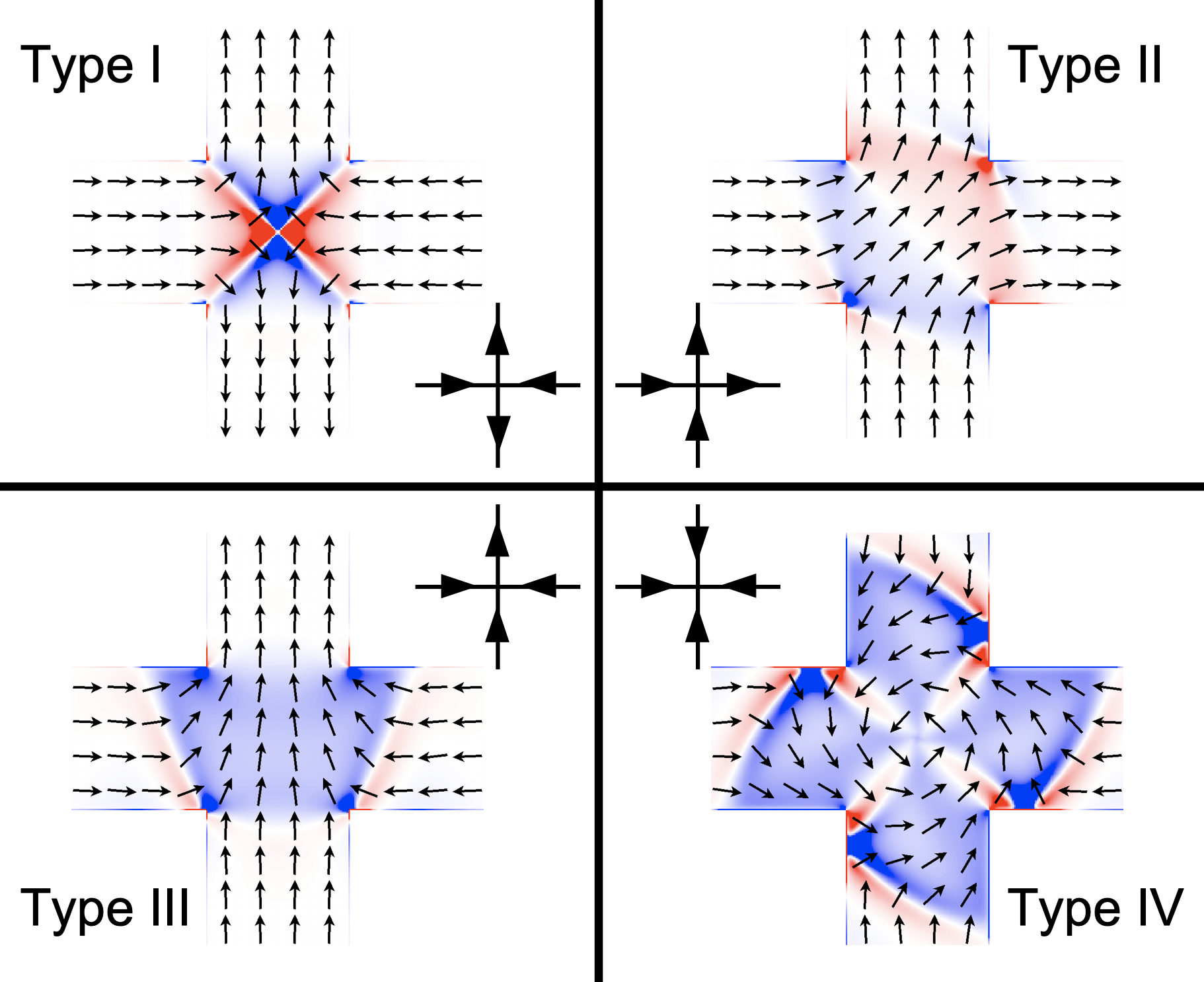}
\caption{\label{micromag} Micromagnetic configurations of type-I, II, III and IV vertices for 200 nm-wide, 20 nm-thick permalloy nanomagnets. Small black arrows represent the direction of magnetization within the nanomagnets, while the blue / red contrast codes for the divergence of the magnetization vector.}
\end{figure}

\section{Micromagnetic simulations}

We performed micromagnetic simulations using the OOMMF code from NIST \cite{OOMMF}. 
The mesh size was set to $2\times 2\times t$ nm$^3$, where $t$ is the thickness of the nanomagnets.
Spontaneous magnetization $M_{s}$ and exchange stiffness $A$ are those of permalloy: $\mu_{0}M_{s}$=1.0053 T, $A$=10 pJ/m, while magnetocrystalline anisotropy is neglected. 
Magnetic moments at the extremities of the nanomagnets are fixed to avoid nonuniform magnetization profiles at the edges.

The micromagnetic configurations for the four types of vertices are reported in Fig. \ref{micromag}. 
Since the nanomagnets are connected, these configurations are similar to magnetic domain walls in many ways.
Type-I vertices have the form of a magnetic antivortex \cite{Mironov2010}, while type-II vertices are almost homogeneously magnetized in (11)-like directions. 
Type-III vertices have the form of a transverse domain wall separating the two horizontal head-to-head nanomagnets. 
Type-IV vertices ressemble a vortex domain wall, with an anticlockwise chirality in Fig. \ref{micromag}.

\begin{figure}
\center
\includegraphics[width=7.5cm]{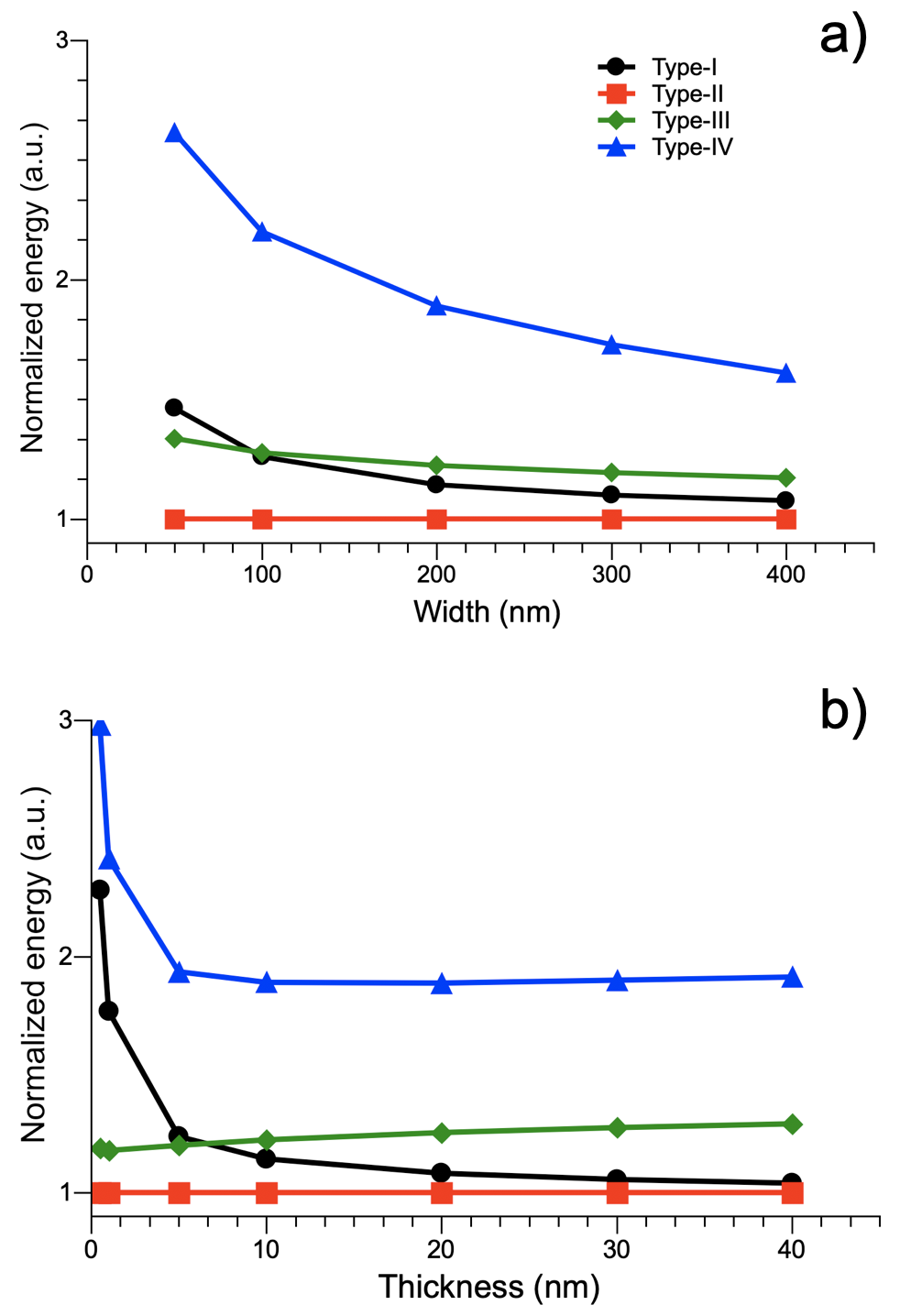}
\caption{\label{curves} Total micromagnetic energy of the four vertex types as a function of the width and thickness of the nanomagnets. Nanomagnets are 10 nm-thick in (a), while they are 200 nm-wide in (b). Type-II vertices have always the lowest energy which is taken here as the reference energy. a.u., arbitrary units.}
\end{figure}

We changed the geometrical parameters of the nanomagnets in a wide range of width [20-400] and thickness [0.5-40], where numbers are in nanometers. 
Fig. \ref{curves} shows how the total micromagnetic energy of each vertex type varies as a function of those parameters. 
These numerical results have two main consequences regarding the type of model that can be potentially accessed experimentally using a square lattice of connected nanomagnets.

First, in the whole range of parameters explored here, type-II vertices have the lowest energy.
Therefore, the ground state configuration consists of independent ferromagnetic straight lines crossing the entire network [see Fig. \ref{vertex}c], like in the KDP model.
This is in sharp contrast with artificial square spin systems made of disconnected nanomagnets in which type-I vertices have always the lowest energy, leading to a twofold degenerate ground state made of local flux-closure configuration [see Fig. \ref{vertex}b]. 
Therefore, a new model can be reached simply by connecting the nanomagnets in a 2d square lattice.
The ground state manifold associated with this model is highly degenerate and the zero-point entropy increases with the lattice size (see next section).

The second consequence deduced from Fig. \ref{curves} is the capability to obtain a realization of the (Lieb) square ice model by using large and thick nanomagnets. 
Indeed, in that case type-I and type-II vertices asymptotitcally tend to have the same energy. 
Clearly, type-II vertices have still the lowest energy, but for large thicknesses, we can envision that sample defects or thermal fluctuations for example will allow the system to select both vertex types equivalently. 
This might offer a unique opportunity to probe the physics of the highly-degenerate manifold of the square ice model.

\begin{figure*}
\center
\includegraphics[width=15cm]{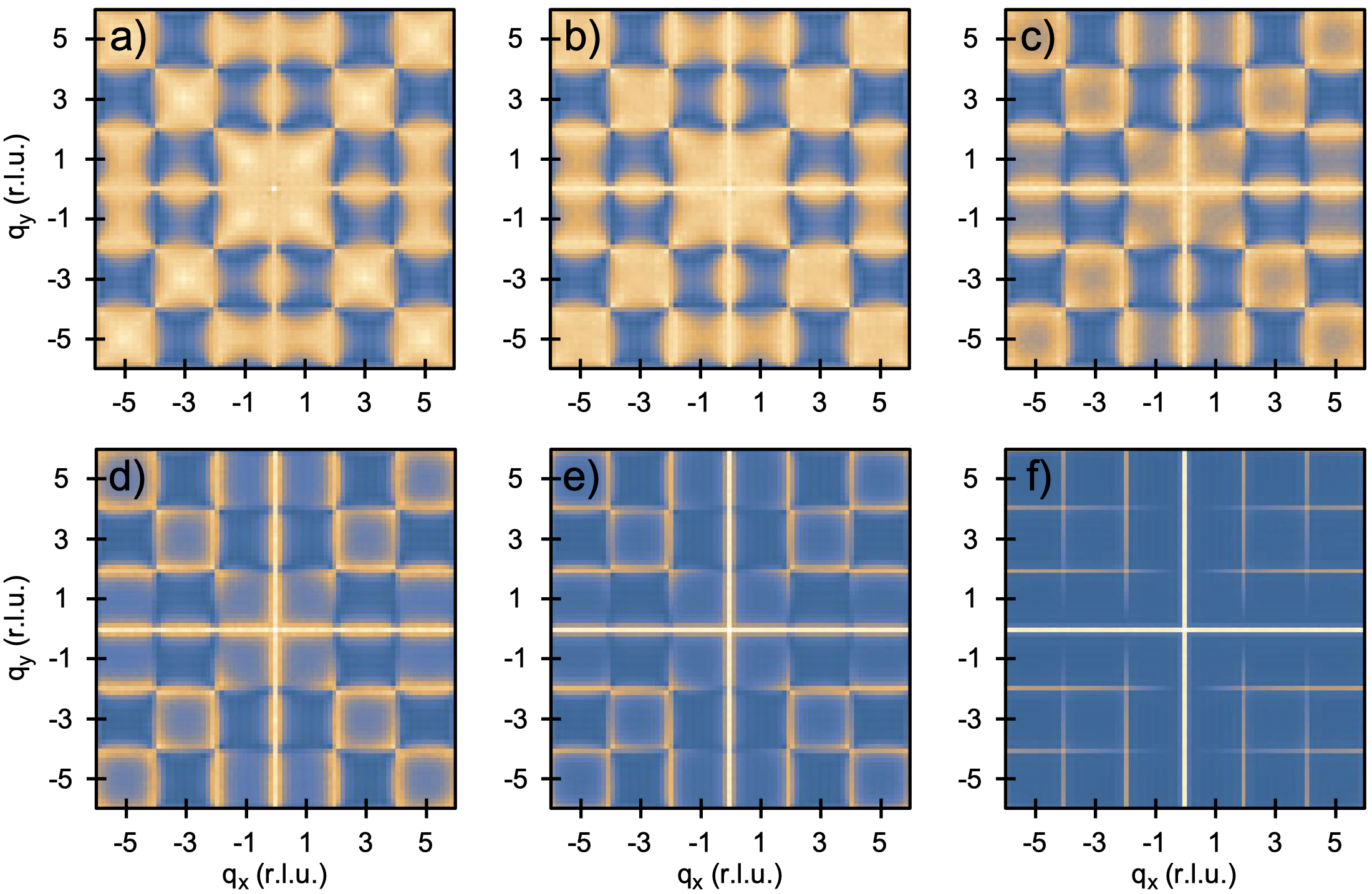}
\caption{\label{MSF} Computed magnetic structure factors at the lowest sampled temperature for $J_2 =$ 1 (a), 1.01 (b), 1.03 (c), 1.05 (d), 1.1 (e) and 1.2 (f). $J_1 = 1$ in all cases. r.l.u., reciprocal lattice unit.}
\end{figure*}

\section{Thermodynamic properties}

We now examine the thermodynamic properties of the spin models associated with the micromagnetic properties described above.
To do so, Monte Carlo simulations were performed using the Hamiltonian
$H= - J_1 \sum_{\langle i  j \rangle} \sigma_i . \sigma_j - J_2 \sum_{\langle \langle i  j \rangle \rangle} \sigma_i . \sigma_j$
, where $\sigma_i$ and $\sigma_j$ are Ising variables on sites $i$ and $j$.
$J_1$ and $J_2$ are positive couplings between orthogonal and collinear nearest neighbors, respectively.
The simulations were done for a $20 \times 20 \times 2$ site lattice with periodic boundary conditions.
A single spin flip algorithm was used to capture the physics that might be observed experimentally, where local magnetization reversal is the only relevant dynamics. 
As a consequence, the simulations suffer from a critical slowing down when approaching the ground state manifold.
At low temperatures, the system freezes, as that would be the case experimentally.
The cooling procedure starts from $T/J_1 = 100$ and ends at $T/J_1 = 0.1$.
10$^4$ modified Monte Carlo steps (mmcs) are used for thermalization \cite{note}.
Measurements follow the thermalization and are computed also with 10$^4$ mmcs.

The spin-spin correlations deduced from the real space configurations are then Fourier transformed for all computed temperatures, leading to a magnetic structure factor.
We define the magnetic structure factor as in neutron scattering experiments, where the spin correlations perpendicular to the diffusion vector are measured \cite{Perrin2016}.
It is composed of a matrix of $81 \times 81$ points covering an area of $\pm 6\pi$ along the $q_x$ and $q_y$ directions in reciprocal space. 

The magnetic structure factors obtained at the lowest sampled temperatures are reported in Figure \ref{MSF} for different values of $J_2$, $J_1$ being fixed to 1.
As expected, when $J_2 = 1$ we find the magnetic structure factor of the square ice, characteristic of an algebraic spin liquid with its associated pinch points \cite{Perrin2016}.
For $J_2$ values larger than 1.1, the magnetic structure factor is drastically different and made of lines spanning across the reciprocal space, consistently with the formation of decoupled ferromagnetic straight lines in real space.
Note that the magnetic structure factor does not show Bragg peaks as the corresponding model does not order and remains liquid-like at low temperature (although subdominant, the total ground state entropy increases with the size of the lattice).
For intermediate $J_2$ values, a continuous transition between the two regimes is observed, where lines coexist with a diffuse background signal. 
Indeed, as $J_2$ approaches $J_1$, the single spin flip dynamics is unable to reach the ground state configuration and the low-temperature physics ressembles the one of the square ice.

\begin{figure}
\center
\includegraphics[width=7.5cm]{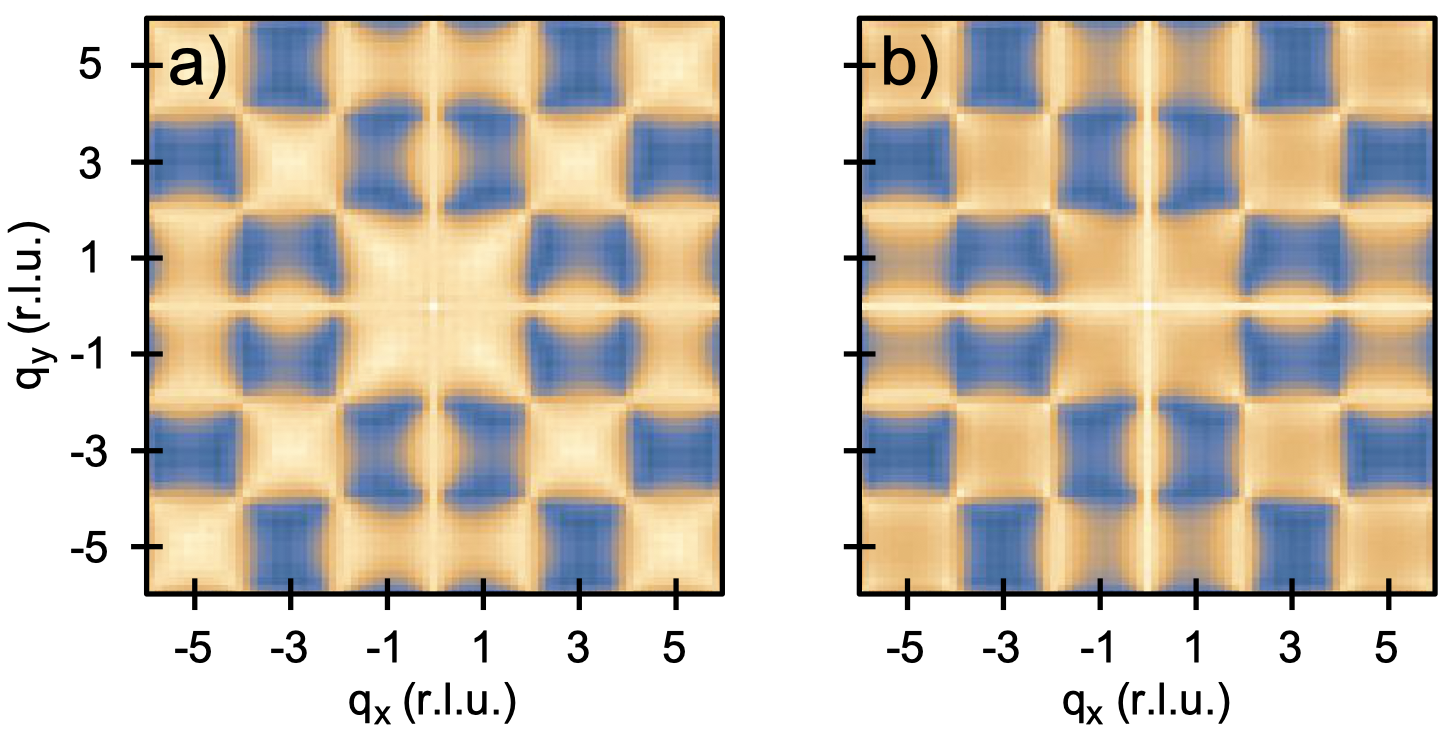}
\caption{\label{comparison} Computed magnetic structure factors at a temperature of $T/J \sim 0.8$ for: (a) $J_2 = 1$ and (b) $J_2 = 1.05$. $J_1 = 1$ in the two cases. r.l.u., reciprocal lattice unit.}
\end{figure}

\section{Discussion}

For coupling strengths that may be reached experimentally, the corresponding low-temperature magnetic structure factor is very similar to the one expected in a pure square ice state.
Considering that:

\noindent - the ground state manifold is difficult to reach in artificial spin systems \cite{Perrin2016},

\noindent - the lowest effective temperature often remains experimentally of the order of $T/J \sim 1$, 

\noindent - the statistics accessible through magnetic imaging techniques is usually poor,

\noindent artificial arrays of connected nanomagnets are expected to be hardly distinguishable from square ice systems.
In fact, even in the Monte Carlo simulations, and after averaging $10^4$ spin configurations, the magnetic structure factors obtained at $T/J \sim 0.8$ are not easy to distinguish at first sight when $J_2 = 1$ and $J_2 = 1.05$ (see Figs. \ref{comparison}a and \ref{comparison}b, respectively).

This comparison suggests that connecting the nanomagnets at the vertex sites of a square lattice might be an easy-to-implement way to approach the physics of the square ice.
Besides, it shows that micromagnetism is potentially a powerful additional degree of freedom that can be used in artificial 2d spin systems to explore degenerate manifolds.
From that prospect, the square geometry offers more flexibility than the kagome geometry as there is more ways to manipulate the micromagnetic texture of a given vertex.

However, we envision that it would be more promising to test our proposal using a thermally active spin system rather than an athermal one subject to a field demagnetization protocol.
The reason is that field driven protocols will likely propagate efficiently magnetic domain walls throughout the lattice, thus favoring the formation of type-II vertices.
Such kinetic effects might be less pronounced in thermally active systems annealed above the Curie point of the consistent material.
We hope that our predictions will stimulate experimental works in that direction.

\bigskip

This work was supported by the Agence Nationale de la Recherche through project through project no. ANR12-BS04-009 'Frustrated'.


\begin{thebibliography}{2}
\bibitem{note}
One modified Monte Carlo step corresponds to a sequence of spin flip attempts such that, on average, each spin is flipped once. This is mandatory to achieve statistical independence. In practice, this corresponds to a number of trials equal to N (number of sites) $\times$ the inverse of the acceptance rate (average value of the spin flip acceptance). At the lowest temperatures, $10^4$ mmcs corresponds typically to $10^8$ spin flip trials.
\end{thebibliography}
\end{document}